\documentclass[twocolumn,showpacs,preprintnumbers,amsmath,amssymb]{revtex4}
\usepackage{graphicx}
\usepackage{dcolumn}
\usepackage{bm}

\newcommand{\bef}{\begin{figure}}
\newcommand{\eef}{\end{figure}}

\newcommand{\be}{\begin{equation}}
\newcommand{\ee}{\end{equation}}
\newcommand{\bea}{\begin{eqnarray}}
\newcommand{\eea}{\end{eqnarray}}
\begin{document}

\title{The width of the rapidity distribution in heavy ion collisions}

\author{Pawan Kumar Netrakanti and Bedangadas Mohanty }
\affiliation{Variable Energy Cyclotron Centre, Kolkata 700064, India}
\date{\today}
\begin{abstract}
We have studied the widths of the rapidity distributions of particles 
produced in nucleus-nucleus collisions at various center of mass energies and 
as a function of centrality at SPS energies. We show that the width 
of the rapidity distribution is sensitive to  
longitudinal flow, velocity of sound in the medium, and rescattering 
of particles. We explore the possibility of  
distinguishing the initial hard scattering regime from 
final state effects by studying the variation in the width of the 
rapidity distribution of the particles with centrality for various 
$p_{T}$ values. 

\end{abstract}
\pacs{25.75.Ld}
\maketitle

The particle production in heavy ion collisions is studied
by measuring the particle density in rapidity ($Y$) or 
in pseudorapidity ($\eta$). The evolution of the pseudorapidity 
density at mid rapidity with beam energy and centrality has 
been one of the main interests 
of study in heavy ion collisions~\cite{dndyexpt}. 
Its scaling with the number of participating nucleons and/or with 
the number of collisions is believed to provide information on the 
dynamics of the particle production~\cite{dndyexpt,dndytheory}. 
The pseudorapidity density at mid rapidity is also related to the 
entropy density~\cite{entropy}. However similar importance has not 
been given to the width of the rapidity distributions of the
particles($\sigma_{Y}$). With the advent of large acceptance 
detectors such as in 
RHIC experiments~\cite{RHIC} and the energy scan being high on the 
agenda of the RHIC program, one can study the evolution 
of the width of the rapidity distribution as a function of beam
energy and  centrality. The width of the rapidity distribution
is believed to be sensitive to the following physics effects:
(a) Final state rescattering~\cite{werner}, hence a $p_{T}$ 
dependence study of the width may help in estimating the value 
of $p_{T}$ that separates the initial hard scattering regime from 
the later stage in heavy ion collisions which is dominated by 
rescattering. (b) The width of the rapidity distribution contains the 
information of longitudinal flow~\cite{ags}. (c) For a given freeze-out 
temperature, the width of the rapidity distribution in the 
Landau hydrodynamical model is found to be sensitive to the velocity 
of sound in the medium~\cite{velo}. 

In this brief report, we first make a compilation of the existing
data on the widths of the rapidity distributions of the particles as a
function of center of mass energy ($\sqrt{s_{NN}}$) 
and centrality. Then we show how longitudinal flow, 
velocity of sound and rescattering effect the width of rapidity 
distribution. Qualitatively the variation in width of the rapidity distribution
with center of mass energy and centrality can be understood on 
the basis of above mentioned processes.

\bef
\begin{center}
\includegraphics[scale=0.35]{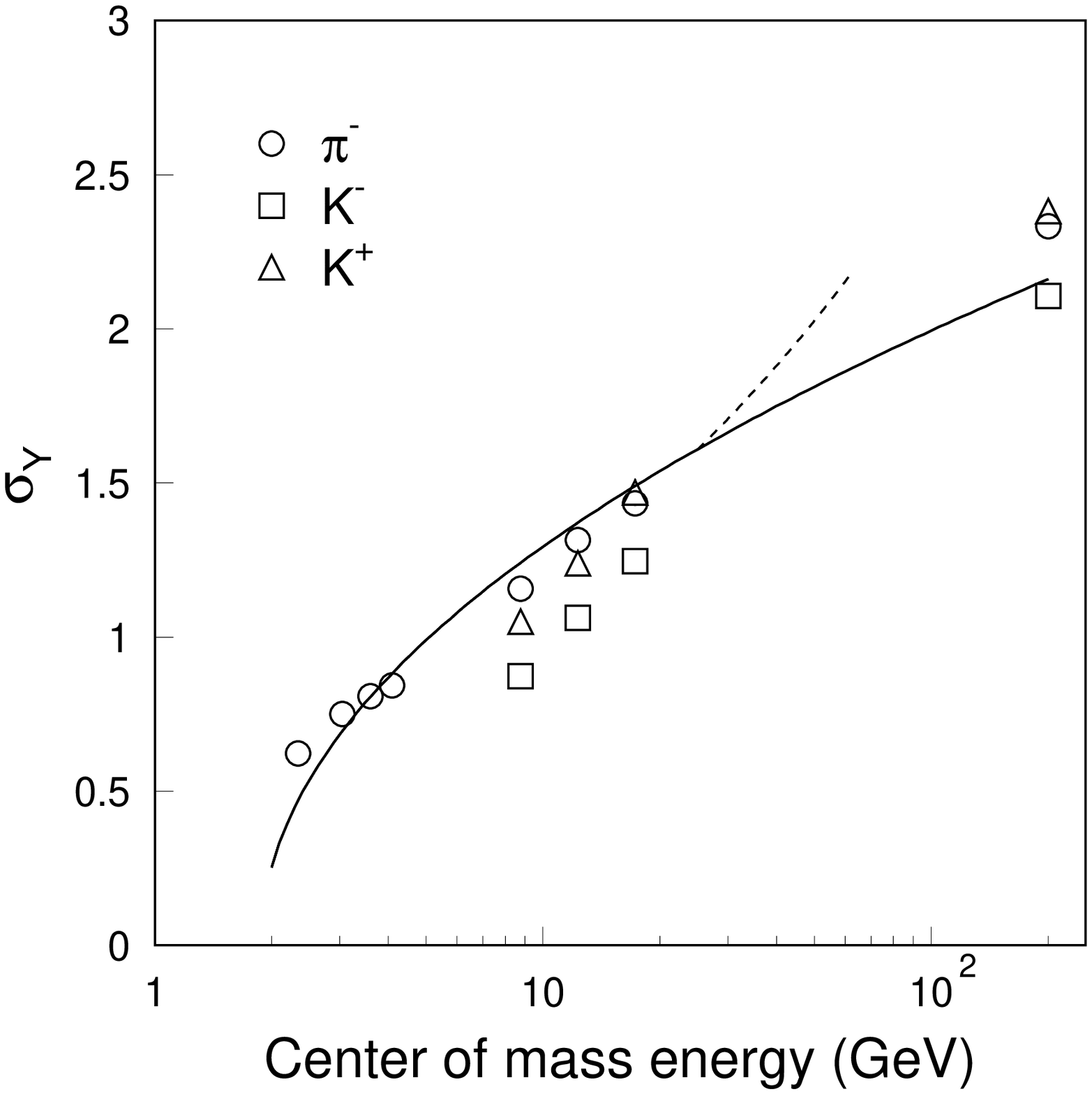}
\caption{Variation of the width of the rapidity distribution for ${\pi^{-}}$, 
$K^{+}$, and $K^{-}$  with center of mass energy. The symbols are for 
nucleus-nucleus collisions and the dashed curve for pions in 
nucleon-nucleon collisions. The solid curve corresponds to 
$\sqrt{\ln(\sqrt{s_{NN}}/2m_{p}})$ from Ref~\cite{Caruthers}. The dashed curve
follows the solid curve up to $\sqrt{s_{NN}}$ = 25 GeV. The errors
on the data point are small and are within the symbol size.}
\label{fig1}
\end{center}
\eef
In Fig.~\ref{fig1} we have plotted 
the width of the rapidity distribution, $\sigma_{Y}$,
for $\pi^{-}$ and $K^{\pm}$ as a function of 
$\sqrt{s_{NN}}$~\cite{ags,sps,rhic2}.
The solid curve corresponds to the theoretical prediction from 
Ref.~\cite{Caruthers} based on the Landau model, developed for studying
the rapidity distribution for pions in $pp$ collisions~\cite{lexus}. 
In this model the width of the rapidity distribution is given as 
$\sqrt{\ln(\sqrt{s_{NN}}/2m_{p}})$. The dashed curve corresponds to the 
experimentally determined width of the rapidity distribution for 
pions produced in $pp$ collisions. 
Where not provided in the references directly, the widths were 
obtained by fitting Gaussian distributions with center at $Y$ = 0 
to the rapidity distributions.
The following observations can be made from the figure: (a) The width of
the rapidity distribution increases with increase in $\sqrt{s_{NN}}$ for 
both nucleon-nucleon and nucleus-nucleus collisions. (b) The width of 
the rapidity distribution for $pp$ collisions deviates from the predictions 
based on the Landau model for $\sqrt{s_{NN}}$ $>$ 25 GeV. 
(c) $\sigma_{K^{+}}$ $>$ $\sigma_{K^{-}}$ in nucleus-nucleus collisions.
This reflects their different interaction cross sections
with other particles in the medium. 
(d) The nucleus-nucleus data for pions shows a similar trend as the 
curve corresponding to $\sqrt{\ln(\sqrt{s_{NN}}/2m_{p}})$. 
$\sigma_{\pi}$ values are comparable to $\sigma_{K^{+}}$ and they are 
higher than $\sigma_{K^{-}}$ values for SPS and RHIC energies. 

\bef
\begin{center}
\includegraphics[scale=0.35]{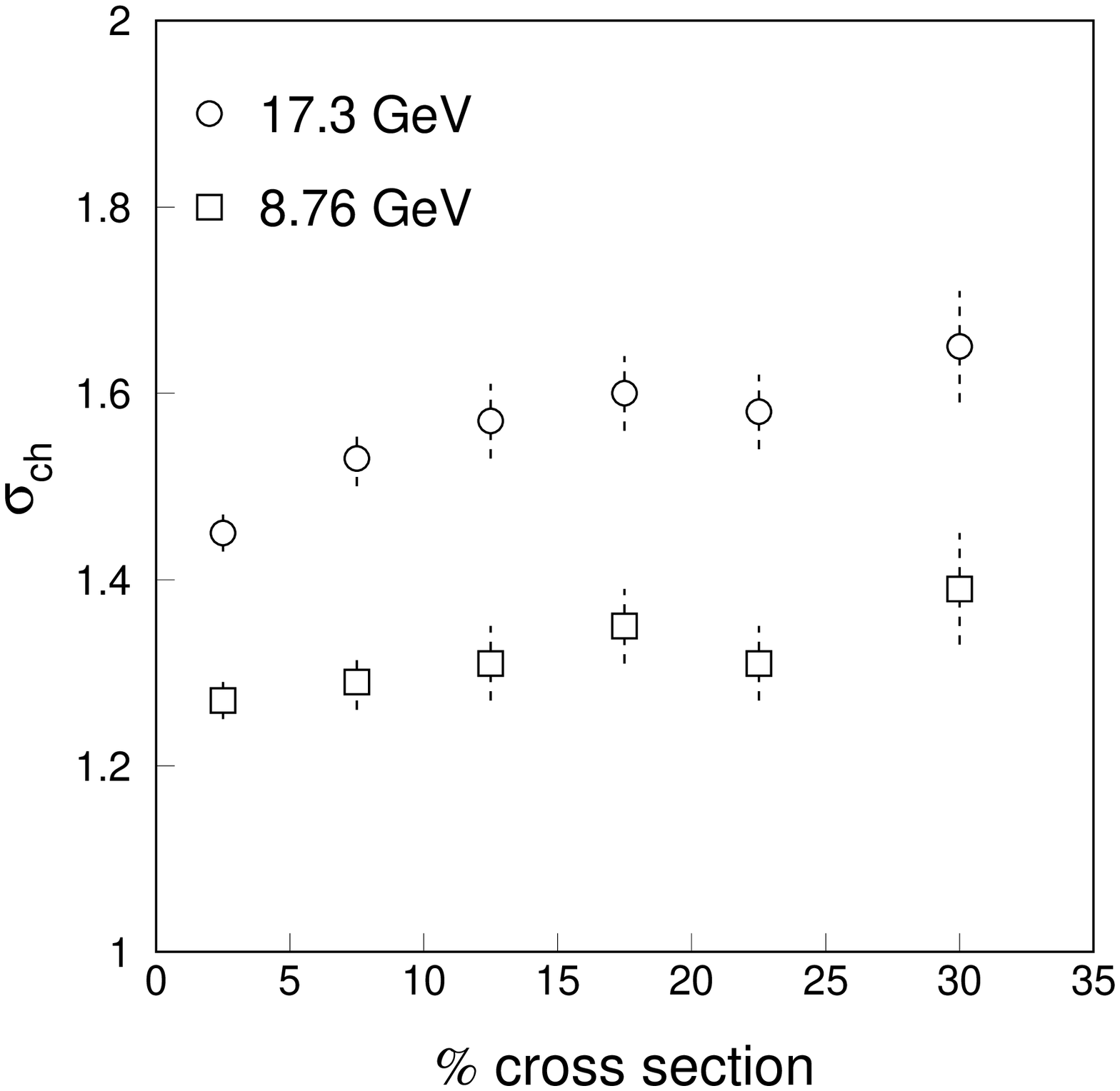}
\caption{Variation of the width of the pseudorapidity distribution of 
charged particles as a function of percentage of cross section for two 
center of mass energies $\sqrt{s_{NN}}$ = 17.3 GeV and 8.76 GeV.}
\label{fig2}
\end{center}
\eef
In Fig.~\ref{fig2} we have plotted the widths of the 
pseudorapidity distributions of the charged 
particles ($\sigma_{ch}$) as a function of \% cross section for 
$\sqrt{s_{NN}}$ = 8.76 and 17.3 GeV~\cite{na50}. We observe that 
$\sigma_{ch}$ increases as we go from central to peripheral 
collisions.

To understand the variation in the width of the rapidity
and the pseudorapidity distribution in Fig.~\ref{fig1} and 
Fig.~\ref{fig2}, we need to know the various physics processes that
effect the width. Here we will qualitatively
discuss how the width of the rapidity distribution is effected
by (a) longitudinal flow, (b) velocity of sound in medium, and (c) initial and 
final state rescattering. 

\bef
\begin{center}
\includegraphics[scale=0.32]{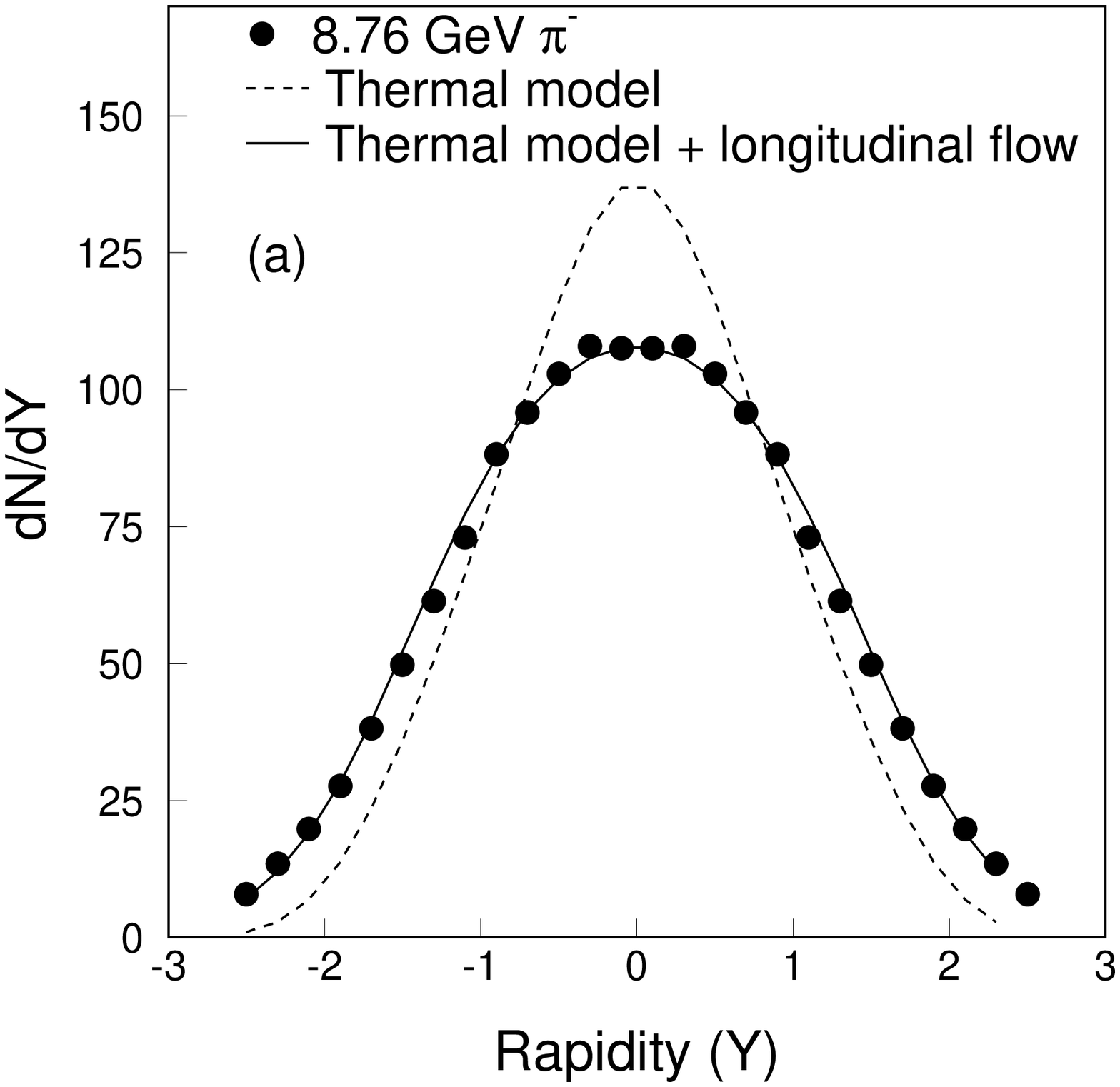}
\includegraphics[scale=0.32]{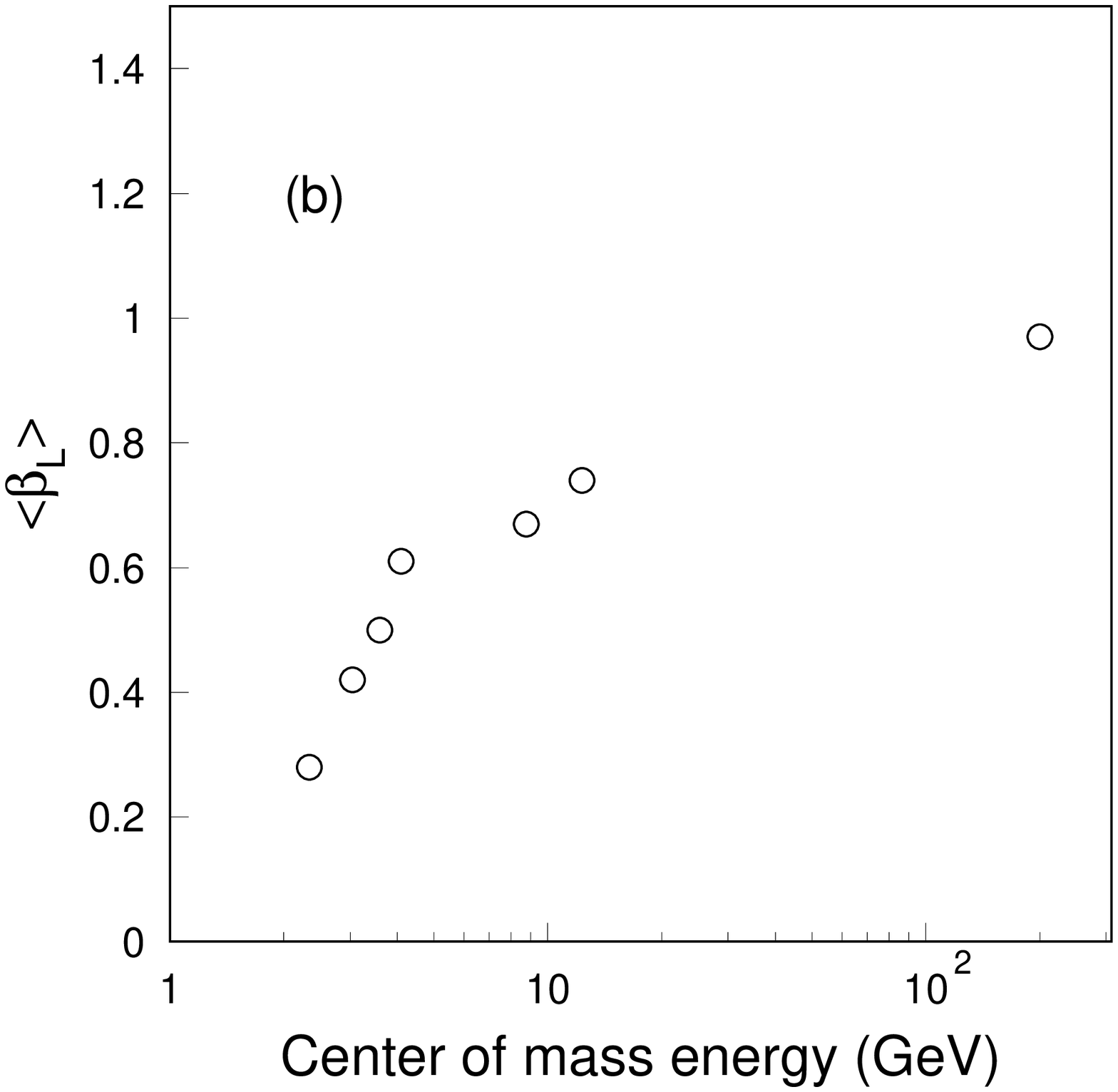}
\caption{(a)Rapidity distribution for ${\pi^{-}}$ at $\sqrt{s_{NN}}$ = 
8.76 GeV~\cite{sps}. Comparison to
a thermal model calculation and a thermal model with longitudinal flow ($\langle \beta_{L} \rangle$ = 0.6) (b)Variation of average longitudinal flow velocity, $\langle \beta_{L}\rangle$, with center of mass energy.}
\label{fig3}
\end{center}
\eef
The rapidity distribution can be used to study the longitudinal 
flow~\cite{ags}. In Fig.~\ref{fig3} we have plotted the rapidity distribution 
of $\pi^{-}$ for $\sqrt{s_{NN}}$ = 8.76 GeV~\cite{sps}. 
Our calculations  from a static 
isotropic thermal emission model where the rapidity density is given as 
\begin{equation}
\frac{dN_{th}}{dY} = AT^3 [ \frac{m^2}{T^2} + \frac{m}{T}\frac{2}{\cosh Y} + \frac{2}{\cosh^{2} Y} ] \times e^{[-(\frac{m}{T}) \cosh Y]}
\end{equation}
is shown by the dashed curve. The temperature $T$ is taken as 
120 MeV~\cite{nuxu}, {\it m} is mass of the pion  
and {\it A} is the normalization constant. We observe that the 
thermal model fails
to explain the width of the rapidity distribution. After including
the longitudinal flow within the ambit of Bjorken hydrodynamics as discussed in
Ref.~\cite{ags,velo}, in the above thermal model, the rapidity distribution 
of pions is found to be well explained (solid curve). 
The rapidity distribution is now given as
\begin{equation}
\frac{dN}{dY} = \int_{-\eta_{max}}^{\eta_{max}} \frac{dN_{th}}{dY} (Y - \eta)~~d \eta
\end{equation}
and the average longitudinal velocity is defined as $\langle \beta_{L} \rangle$
= tanh$(\eta_{max}/2)$.
$\langle \beta_{L} \rangle$ = 0.6 is found to explain the pion 
data at $\sqrt{s_{NN}}$ = 8.76 GeV as shown by the solid curve in 
Fig.~\ref{fig3}(a).

We fitted the rapidity distributions for pions 
from $\sqrt{s_{NN}}$ = 2 to 200 GeV
to results from the thermal model with longitudinal flow
to obtain the $\langle \beta_{L} \rangle$. Variation of 
$\langle \beta_{L} \rangle$ with $\sqrt{s_{NN}}$ is shown in 
Fig.~\ref{fig3}(b). We find the average longitudinal velocity for pions 
approach a value of 1 at RHIC from a value of 0.3 at AGS energies.
This is indicative of the higher degree of nuclear transparency attained
at RHIC compared to SPS or AGS.
It is observed that the results in Fig.~\ref{fig3}(b) show qualitatively a 
similar trend with $\sqrt{s_{NN}}$ as seen for the $\sigma_{Y}$ for pions 
in Fig.~\ref{fig1}. 
This indicates that the collective behaviour and the final state interactions
of the produced particles in nucleus-nucleus collisions play an important role
in determining the width of the rapidity distribution.

\bef
\begin{center}
\includegraphics[scale=0.35]{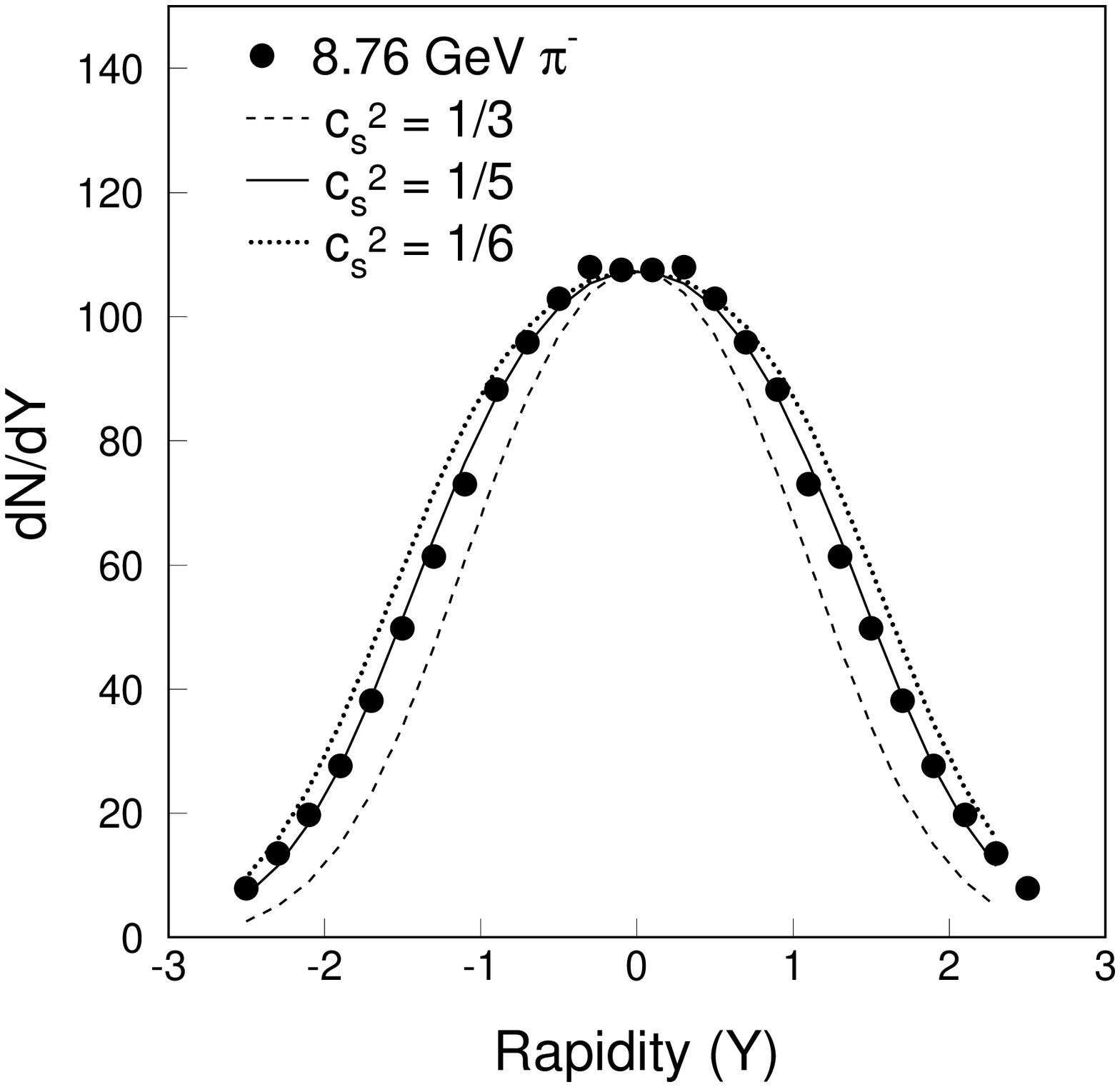}
\caption{Rapidity distribution of $\pi^{-}$ at center
of mass energy 8.76 GeV compared to the rapidity spectra obtained from
the Landau hydrodynamical model for $c_{s}^2$ = 0.166, 0.2 and 0.33.}
\label{fig4}
\end{center}
\eef
The width of the rapidity distribution is sensitive to the velocity of
sound in the medium formed at freeze-out~\cite{velo}. 
Fig.~\ref{fig4} shows the rapidity distribution of pions 
at $\sqrt{s_{NN}}$ = 8.76 GeV compared to rapidity distribution 
obtained for various values of velocity of sound using  
Landau hydrodynamics. Within the ambit of Landau hydrodynamics one
can show, with certain assumptions~\cite{velo}, that the rapidity 
distribution has the form
\begin{equation}
\frac{dN}{dY}\,\sim\,Const.\frac{\exp(-\frac{Y^2}{2\sigma^2})}
{\sqrt{2\pi\sigma^2}} \,
\label{eqprime}
\end{equation}
where $\sigma=2\omega_f/(1-c_s^2)$, $\omega_{f}$ = ln$(T_f/T_0)$, 
$T_f$ is the freeze-out temperature, $T_0$ is the initial temperature,
$c_s$ is the velocity of sound in the medium.
We observe that for a $T_f$ = 120 MeV~\cite{nuxu} and $T_0$ = 230 MeV 
(obtained from the total multiplicity~\cite{entropy}),
 $c_{s}^2$ = 1/5 explains the data very well. 
A $c_{s}^2$ value of 1/6 overpredicts the data and a $c_{s}^2$ value
of 1/3 (ideal gas) underpredicts the data. The $\chi^2$ values 
for the distributions with $c_{s}^2$  = 1/3, 1/5 and 1/6 are
51, 1, 7  respectively. 
This shows that the width of the rapidity distribution of data is 
sensitive to the parameter  $c_{s}^2$ representing the velocity of 
sound in the medium in the above model. 
It may be mentioned that the results are sensitive to the choice of 
initial and freeze-out temperatures also~\cite{velo}.

Our analysis of the rapidity distributions of pions, kaons and protons at 
AGS and SPS energies all reveal the same value of $c_{s}^2$ $\sim$ 1/5 
which explains the data. This may indicate some kind of universality of the 
matter formed at the freeze-out stage. It may be mentioned that the 
value of $c_{s}^2$ $\sim$ 1/5 has been found to be a  characteristic
value of the speed of sound for a gas of hadrons~\cite{shuryak}.
$c_{s}^2$ $\sim$ 1/5 indicates that the expansion of the system is slower
than that in an ideal gas scenario ($c_{s}^2$ = 1/3). Thus the 
system formed in heavy ion collisions gets more time to interact and 
to reach thermal equilibrium.

\bef
\begin{center}
\includegraphics[scale=0.35]{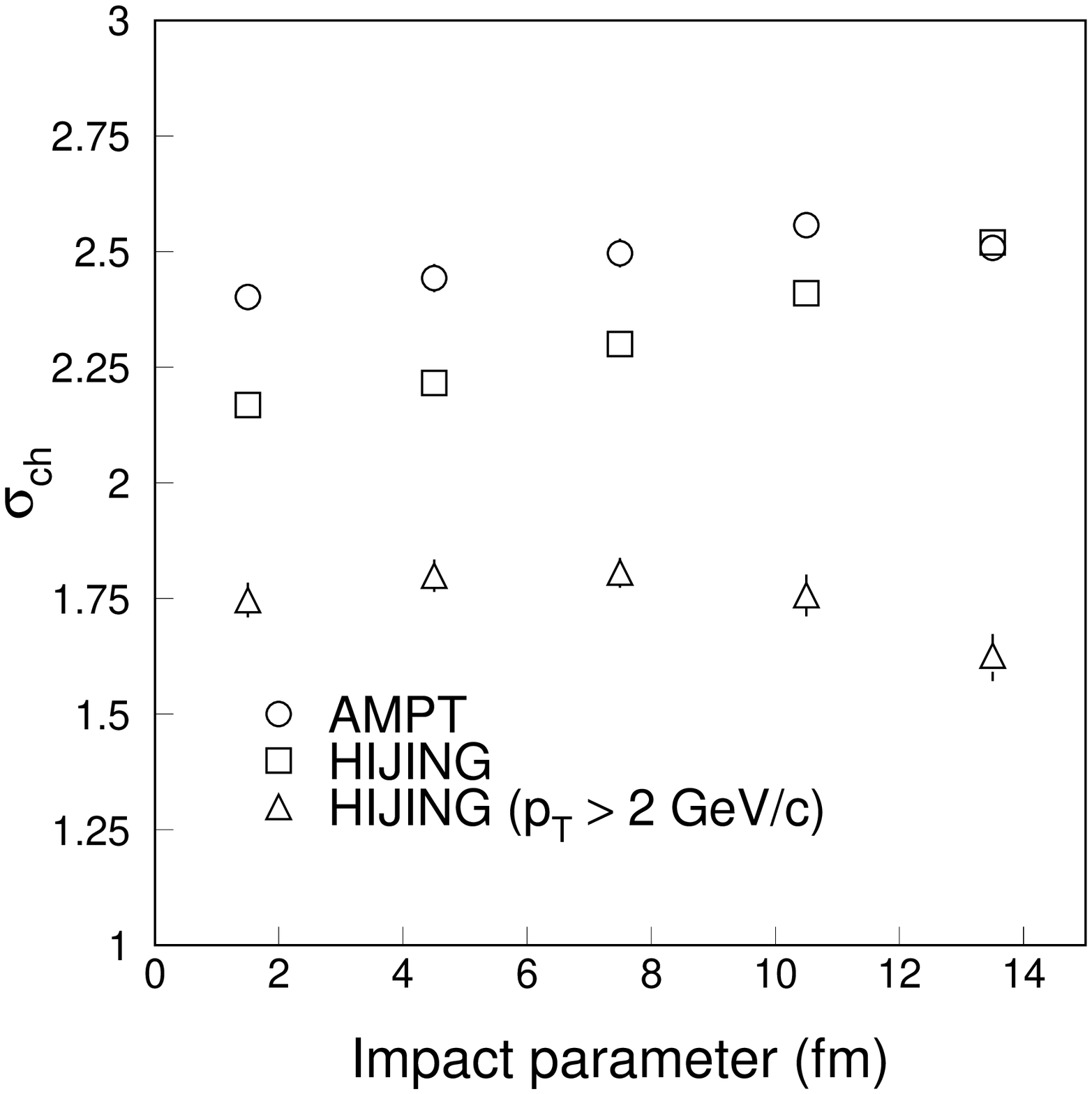}
\caption{Variation in the width of the rapidity distribution as a function
of impact parameter for $\sqrt{s_{NN}}$ = 62.4 GeV for Au on Au 
collisions from AMPT and HIJING models. Also shown width of rapidity 
distribution from HIJING for particles having $p_{T}$ $>$ 2 GeV/{\it c}.  }
\label{fig5}
\end{center}
\eef
Sensitivity of the width of the rapidity distribution to rescattering 
effects is studied here by using A Multi Phase Transport (AMPT) 
model~\cite{ampt}. The AMPT model includes both initial partonic 
and final hadronic interactions. It uses the input parton distribution
from the HIJING model~\cite{hijing}. We found that $\sigma_{ch}$ 
obtained from the AMPT model increases with $\sqrt{s_{NN}}$. 
The values at $\sqrt{s_{NN}}$ = 17.3, 62.4, 130 and 200 GeV 
are 1.57, 2.4, 2.6 and 2.8 respectively for the impact parameter 
range from 0 to 3 fm in Au+Au collisions. In Fig.~\ref{fig5} 
the widths of the rapidity distributions for charged
particles is plotted as a function of impact parameter for 
$\sqrt{s_{NN}}$ = 62.4 GeV for Au on Au collisions using AMPT 
and HIJING models. We observe that $\sigma_{ch}$ increases as we go
towards higher values of impact parameter for both AMPT and HIJING
models. We also observe such a trend at lower and higher 
center of mass energies. This trend is qualitatively similar 
to that observed in the data shown in Fig.~\ref{fig2}. The 
variation in the width is smaller for AMPT than for
HIJING. This indicates that final state rescattering has an 
effect on the width of the rapidity distribution. 

Also shown in Fig.~\ref{fig5} is the variation of the $\sigma_{ch}$ from 
HIJING for the particles with $p_{T}$ $>$ 2 GeV/{\it c} as a function of 
impact parameter. 
We observe that for particles having $p_{T}$ $>$ 2 GeV/{\it c}, 
$\sigma_{ch}$  decreases as we go higher in collision impact parameter. 
Qualitatively one can think of the following picture,
rescattering leads to more isotropic momentum distributions (for example
radial flow in a hydrodynamical picture) and hence will lead to 
narrower rapidity distributions. Particles with very high transverse 
momentum which are basically coming from the initial state will not exhibit 
such isotropy in the momentum distribution. 
The width of their rapidity distribution is expected to show a different 
variation with centrality.
$p_{T}$ $>$ 2 GeV/{\it c} was chosen for this study 
as RHIC results on elliptic flow show that hydrodynamical 
calculations agree with the data for  $p_{T}$ $<$ 2 GeV/{\it c}~\cite{star_flow}.
Thus studying  $\sigma_{ch}$ as a function of centrality 
for various $p_{T}$ ranges may indicate the possibility of 
finding a value of transverse momentum at which the initial 
hard scattering stage can be distinguished from the later final 
state rescattering.

In summary, we have found that the width of the rapidity distribution
of particles increases with increase in $\sqrt{s_{NN}}$ 
in heavy ion collisions. 
$\sigma_{K^{+}}$ $>$ 
$\sigma_{K^{-}}$ reflects the different interaction cross sections
 of oppositely charged kaons 
with other particles in the medium. It has also been observed in the data
that the width of the rapidity distribution increases with increase in
impact parameter.  We have shown that the width of the rapidity 
distribution can be effected by longitudinal flow, velocity of
sound, and rescattering. 
Longitudinal flow is found to 
increase with increase in $\sqrt{s_{NN}}$. 
Qualitatively the variation of average longitudinal flow velocity with 
$\sqrt{s_{NN}}$ shows a similar trend as $\sigma_{Y}$ for pions 
with $\sqrt{s_{NN}}$.
For a given freeze-out and
initial temperature the width of the rapidity distribution is sensitive
to the velocity of sound in the medium. 
A value of  $c_{s}^2$ $\sim$ 1/5 is able
to explain the rapidity distribution of particles at AGS and SPS
energies. This indicates, the expansion of the system is slower than 
compared to in an ideal gas scenario ($c_{s}^2$ = 1/3). Thus the 
system formed in heavy ion collisions gets more time to interact and 
to reach thermal equilibrium. The results from a multi-phase transport 
model qualitatively show a similar dependence of the width of the 
rapidity distribution on the impact parameter as does the data. 
We have also studied the possible $p_{T}$ dependence of the width 
of the rapidity distribution of charged particles as a function 
of impact parameter to get an idea about initial hard scattering 
and final state effects.


\normalsize
\begin{thebibliography}{99}
\bibitem{dndyexpt} 
                   B.B. Back {\it et al.} (PHOBOS Collaboration),
	           Phys. Rev. Lett {\bf 85}, 3100 (2000);
                   K. Adcox {\it et al.} (PHENIX Collaboration),
                   Phys. Rev. Lett {\bf 86}, 3500 (2001);
                   P.K. Netrakanti and B. Mohanty,
                   Phys. Rev. C {\bf 70}, 027901 (2004).
\bibitem{dndytheory} 
                   X-N. Wang and M.~Gyulassy, 
	           Phys. Rev. Lett {\bf 86}, 3496 (2001);
                   D. Kharzeev and M.~Nardi, 
                   Phys. Lett. B507 (2001) 121;
                   D. Kharzeev and E.~Levin, 
                   Phys. Lett. B523 (2001) 79;
                   A. Capella and D. Sousa,
                   Phys. Lett. B {\bf 511} (2001) 185.
\bibitem{entropy}  
                   B. Mohanty, J. Alam and T.K. Nayak,
                   Phys. Rev. C {\bf 67}, 024904 (2003).

\bibitem{RHIC} 
	           Proceedings of Quark Matter 2004 
	           [Jour. of Phys. G 30 (2004)].
\bibitem{werner} 
                   J. Aichelin and K. Werner, 
	           Phys. Lett. B {\bf 300}, 158 (1993).

\bibitem{ags} 
                   J. L. Klay {\it et al.}, (E895 Collaboration),
	           Phys. Rev. Lett {\bf 88}, 102301 (2002);
                   Phys. Rev. C {\bf 68}, 054905 (2003).                   
\bibitem{velo} 
                   B. Mohanty and J. Alam,
                   Phys. Rev. C {\bf 68}, 064903 (2003).
\bibitem{sps} 
                   S. V. Afanasiev {\it et al.}, (NA49 Collaboration),
                   Phys. Rev. C {\bf 66}, 054902 (2002).
\bibitem{rhic2} 
                   I.G. Bearden {\it et al.}, (BRAHMS Collaboration),
                   nucl-ex/0403050.
\bibitem{Caruthers} P. Carruthers and M. Duong-van, 
                    Phys. Lett. B {\bf 41}, 597 (1972);
                    Phys. Rev. D {\bf 8}, 859 (1973).
\bibitem{lexus} 
                   S. Jeon and J. I. Kapusta,
                   Phys. Rev. C {\bf 63}, 011901 (2001).
\bibitem{na50}     
                   M.C. Abreu {\it et al.}, (NA50 Collaboration), 
                   Phys. Lett. B {\bf 530}, 43 (2002).
\bibitem{nuxu}     
                   N. Xu and M. Kaneta,
                   Nucl. Phys. A {\bf 698}, 306 (2002). 

\bibitem{shuryak}  
	           D. Teaney, J. Lauret and E.V. Shuryak,
                   nucl-th/0110037.

\bibitem{ampt} 
               B. Zhang, C.M.~Ko, B.-A. Li, and Z.~Lin, 
               Phys. Rev. C {\bf 61}, 067901 (2001). 
\bibitem{hijing} 
               X-N. Wang and M.~Gyulassy, 
               Phys. Rev. D {\bf 44}, 3501 (1991); 
               Comput. Phys. Commun. {\bf 83}, 307 (1994).

\bibitem{star_flow} 
                   K. H. Ackermann {\it et al.}, (STAR Collaboration),
	           Phys. Rev. Lett {\bf 86}, 402 (2001).

\end{thebibliography}
\end{document}